\documentclass[12pt]{article}
\usepackage{amssymb}
\usepackage{graphics}
\usepackage{graphicx}
\begin{document}
\title{Can One Make Any Crash Prediction in Finance Using the Local Hurst Exponent Idea?}
\author{D. Grech
\\
  Institute of Theoretical Physics\\
   University of Wroc\l aw, Pl.M.Borna 9,
\\ 50-204 Wroc\l aw, Poland \\
 e-mail: dgrech@ift.uni.wroc.pl
\\
and \\ Z. Mazur
\\
Institute of Experimental Physics\\
 University of Wroc\l aw, Pl.M.Borna 9,
  \\ 50-204 Wroc\l aw, Poland}
\date{}
\maketitle

\begin{abstract}
We apply the Hurst exponent idea for
investigation of DJIA index time-series data. The  behavior of the
local Hurst exponent prior to drastic changes in financial series
signal is analyzed. The optimal length of the time-window over
which this exponent can be calculated in order to make some
meaningful predictions is discussed. Our prediction hypothesis is
verified with examples of '29 and '87 crashes, as well as with
more recent phenomena in stock market from the period 1995-2003.
Some interesting agreements are found.
\end{abstract}
%\begin{keyword}
%%Econophysics, Time series, Correlations, Brownian motion, Scaling laws, Hurst exponent
%\end{keyword}

\section{Introduction}

 Financial markets are nonlinear, complex and open dynamical
systems described by enormous number of free, mostly unknown
parameters. Many of these parameters are exterior for the market,
what makes the full description of the system even more
complicated. Exterior  parameters are completely out of control by
investors who are the part of financial system. These parameters
have usually random origin connected with e.g. political
disturbances, terrorist attacks, bankruptcies of leading
companies, wars, etc. There are also investor dependent interior
parameters driving the market even in the absence of other, not
expected exterior phenomena. The full recognition of such interior
parameters
is a big challenge for economists and econophysicists.\\
All these circumstances mean that detailed time evolution of a
complex financial system is unpredicted. However is the situation
so hopeless? Even if we can not predict the detailed evolution
scenario, we might be able to say something else about the system.
We have learned  from statistical mechanics that one does not have
to know where a particular particle of the mechanical system will
be a second or few from now, in order to find equation of state of
this system. The latter is quite sufficient in practical
applications giving us the whole available information about the
macroscopic parameters, like e.g. pressure or temperature, and it
reflects all detailed microscopic, directly inaccessible
information. In many cases this knowledge is sufficient to
indicate direction in which the system is
evolving.\\
In this article we rise the question if it is possible to find
some macroscopic parameters in financial market which would play
the role of a macroscopic indicator of complex, interior stock
dynamics. In particular, we would like this parameter to be able
to predict that crashes or other drastic changes in the market
signal are coming soon. The task looks hopeless if one assumes
following Fama [1] that investors destroy information while using
it, so that others can not use this information again. In this
spirit market should not exhibit any correlations between returns
$ r_t$ at time $t$ and $r_{t+\tau}$ at time $t+\tau$, where
\begin{equation}
r_t = \ln{\frac{S_{t+1}}{S_t}}
\end{equation}
and $S_t$ is the price of a given stock at time $t$.\\
It was widely believed for a long time that price changes follow
an independent, zero mean, Gaussian process. However, deviations
from this simple scenario have been observed in the financial
signal in the last few years. Empirical work showed that the
distribution function $P_r(r>x)$ has tails obeying power law
relation $P_r(r>x)\sim x^{-\alpha}$ [2-4], contrary to normal
Gaussian  distribution. Also the autocorrelation function of the
{\it absolute value} of price changes shows long-range persistence
$\langle|r_t||r_{t+\tau}|\rangle\sim\tau^{-\mu}$, with $\mu\approx
0.3$ [5-7]. A sort of long memory correlations in
financial signal itself has also been revealed [8-12].\\
Due to large liquidity of currency exchange markets, they seem to
be a natural subject to explore the existence of such correlations
between returns. This has been done for major world and European
currencies in [13-15]. It has also been proven that complexity of
a financial market is not limited to the statistical behavior of
each financial time series forming it but follows from the statics
and dynamics of correlations existing between various stocks in the market [16].\\
The proven existence of correlations in financial time series
reveals the possibility to apply some global macroscopic approach
to see them. In this paper we address the possibility of searching
for correlations between subsequent returns in financial series
with the use of Detrended Fluctuation Analysis (DFA)[17], applied
for the first time in finances in [18-20]. Our philosophy is
described below.

\section{The Search for the Market State with the Use of Local Hurst Exponent}

It is well known that the Hurst $\alpha$-exponent [21,22],
extracted from the time series according to DFA method, measures
the level of persistency in the given signal. The value $\alpha
\neq 1/2$ implies the existence of long-range correlations and
corresponds to  so called fractional Brownian motion [23]. In
particular, for $\alpha>1/2$ there is persistence, and for
$\alpha<1/2$ 'antipersistence' in the series signal. We expect
that dramatic changes in financial signal should be preceded by
excitation state of the market (nervousness), what in turn is
reflected by the shape of subsequent daily changes in the signal.
These changes should become less correlated just before the
dramatic breakdown in the signal trend. Contrary, when the trend
in the market is strong and well determined, an increasing
(decreasing) value of market index in immediate past makes also an
increasing (decreasing) signal in the immediate future more
probable [24]. In other words, one should observe some long-range
correlations in returns and consequently higher $\alpha$ values
for strong, long lasting trends (increasing or decreasing ones),
and significant drop in $\alpha$-exponent value if the trend is
going
to change dramatically its direction in very near future.\\
To check this hypothesis we used DFA technique, rather than other
available methods like spectral analysis or rescaled range
analysis, to extract $\alpha$-exponent. This is because DFA method
avoids detection of long-range correlations being an artefact of
nonstationarity  of time series [25]. Then some applications were
performed for
Dow Jones Industrial Average (DJIA) daily closure signal.\\
For completeness let us briefly remind the main steps of DFA
analysis:
\begin{enumerate}

\item {The time series of random one variable sequence $x(t)$ of
length $N$ is divided into $N/\tau$ non-overlapping boxes of equal
size $\tau$. The time variable is discrete and evolves by a single
unit between $t=1$ ($x(t=1)\equiv x_1$) and $t=N$  $(x(t=N)\equiv
x_N)$. Thus each box contains $\tau$ points and $N/\tau$ is
integer.} \item {The linear approximation of the trend in each
$\tau$-size box is found as  $y_\tau(t) = a_\tau t + b_\tau$, with
$a_\tau, b_\tau$ some box dependent constants.} \item{ In each
$\tau$-size box one defines the 'detrended walk' $x_\tau (t) =
x(t) - y_\tau(t)$ as the difference between the original series
$x(t)$ and the local trend $y_\tau(t)$.} \item{ One calculates the
variance about the detrended walk for each box:
\begin{equation}
F^2_i(\tau) = \frac{1}{\tau}\sum\limits_{t\in i-th \, box}(x(t) -
y_\tau(t))^2 \equiv \frac{1}{\tau}\sum\limits_{t\in i-th \,
box}x_\tau^2(t)
\end{equation}
 and the average of these variances over all $N/\tau$ boxes of size $\tau$:
\begin{equation}
\langle F^2(\tau)\rangle =
\frac{\tau}{N}\sum\limits^{N/\tau}_{i=1}F^2_i(\tau)
\end{equation}
} \item{A power law behavior is expected:
\begin{equation}
\langle F^2(\tau)\rangle \sim \tau^{2\alpha}
\end{equation}
from which $\alpha$-exponent can be extracted from log-log linear
fit.}
\end{enumerate}
If one wants to use $\alpha$-exponent to measure the strength of
local correlations in financial time series, one has to use the
{\it local} $\alpha$-exponent idea [17, 18]. For a given trading
day $t=i$, the corresponding $\alpha_i$-exponent value will be
calculated according to Eq.(4) in the period $\langle
i-N+1,i\rangle$ of length $N$, called an observation box or
time-window. In order to cover the whole time-window length $N$
with $\tau$-size boxes, we put the last box contributing to Eq.
(3) in the period $\langle i-N+1, i - [ \frac{N}{\tau}]\tau+1
\rangle$, where $[.]$ means the integer part. This box partly
overlaps the preceding one but such overlapping does not modify
the local $\alpha$. Moving the time-window every one session, one
is able to reproduce the history of $\alpha$ changes in time. Let
us notice that only the past signal of financial series, not
earlier than $N$ sessions before a given
trading day $t$, contributes to local $\alpha$ value.\\
It is not surprising that local $\alpha$-exponent at given moment
$t$ depends on the time-window length $N$. It is seen in Fig.1
where we present the example of local $\alpha$-exponent plots for
chosen $400$ trading day period  of DJIA daily closure values.
Three different choices: $N=210$, $N=350$ and $N=420$ have been
here made to calculate local $\alpha$. One may notice that plots
for $N=350$ and $N=420$ coincides very well with each other, while
$N=210$ plot shows already some deviations from other two. It
exhibits slightly higher local $\alpha$ values in the first half
of discussed period than in the second half. The question arises,
whether this is a real effect outside the statistical
uncertainties range and what the optimal
time-window length $N$ should be chosen  for further discussion.\\
The choice of $N$ seems to be a matter of intuition, statistics
and economic regards. If $N$ is too large, $\alpha$-exponent loses
its locality and may not 'see' correlations whose characteristic
range, say $\rho$, is much smaller than the window length
($\rho\ll N$). This is the main reason why $\alpha$ evolution
becomes more smooth if $N$ increases. On the other hand, it has
been proven that standard deviation $\Delta\alpha$ in the local
$\alpha$ value, caused by finite-size time-series effects on
long-range correlation, is of order [26]:
\begin{equation}
\Delta\alpha\sim(\frac{\tau}{N})^{1/2}
\end{equation}

Thus, we are stuck with huge statistical
uncertainty with decreasing $N$, what efficiently destroys all predictions.\\
One has to find then a golden point (golden area) where two above
requirements meet together, i.e. where $\Delta\alpha$ is
sufficiently small, and $N$ is not too large. For economic reason
we suggest that $N$ should not exceed one trading year ($N=240$)
in the case of financial series with daily closure signal. This is
to avoid probable seasonal periodicity in supply and demand in the
market, what would introduce artificial contributions to
investigated correlations. Besides, in our opinion, $\alpha$ loses
a sens
of the local value for $N>240$. In order to stay within the scaling
range of Eq. (4) we used the box size $5<\tau<N/5$.\\
To find the optimal $N$ we investigated how far the standard
deviation $\Delta\alpha$ changes with $N$ for $N\leq 300$. This
has been done for different subseries of DJIA signal, taking $500$
samples for each value of $N$. The results are collected in Fig.
2, where $\Delta\alpha(N)$ is displayed together with the
respective percentage uncertainty $\Delta\alpha/\langle
\alpha\rangle$. As one would expect from Eq. (5),
$\Delta\alpha(N)$ slowly decreases on the average with $N$, but
some departs from this rule are observed. The first local minimum
below $N=240$ appears at $N=215$. Although this minimum is not
strong, we decided to make further estimations of local $\alpha$
for DJIA series using this particular time-window length. Our
choice corresponds to about ten months trading period. It is worth
to observe that the obtained statistical uncertainty $\sim 9\%$
agrees well with that calculated in [18] for a monetary market. We
have also checked that, for $N<200$, there exists a substantial
drop of linear regression correlation coefficient $R^2$, what
additionally justifies our choice. Finally, we found that the
local $\alpha$ value is not sensitive to changes in $N$ not
exceeding $10\%$, so that any choice $190\leq N \leq 230$
reproduces qualitatively and quantitatively very similar results.

\section{DJIA Signal Tested with DFA Method - Applications}

Now we proceed to analyze  the local value of $\alpha$-exponent
estimated with the DFA technique for daily closure DJIA signal.
The whole history of this signal is collected in Fig.3. We have
focused attention on the most important events in the market
history. These are: crashes in September 1929, October 1987, July
1998, and the current situation in last three years. All data were taken from [27].\\
First let us consider the '29 crash shown in details in Fig.4. The
DJIA signal had been in clear increasing mode for about $4$ months
- from session $\sharp 9760$ up to the session $\sharp 9840$
(3.09.1929) when the trend in market index changed its direction.
The crash took place 40  days  later. The detailed
$\alpha$-exponent structure of this period is shown in Fig.4a,
where dots denote respective local $\alpha$ values calculated
session by session from the last $N=215$ index values. To see more
effectively the evolution of local $\alpha$, the moving average
value $\alpha_{m5}$ of $\alpha$-exponent taken from five last
sessions (one trading week) has been drawn in Fig.4b. A very clear
decreasing trend in local $\alpha$ is visible. It had started
about one month before the DJIA index reached its maximal value on
3.09.1929, but did not terminate with this date. Small corrections
in local $\alpha$ around  the session $\sharp 9840$ can be
explained as a result of change in DJIA index trend  at that time.
The $\alpha$-exponent has reached a deep and clearly seen local
minimum $\alpha\sim 0.45 $ two weeks before the crash, contrary to
very high value $\alpha\sim 0.65$ from which it started several
months earlier. The statistical uncertainty in the difference
between initial ($\alpha_0$) and final ($\alpha_f$) values is at
most $(\Delta\alpha_0/\alpha_0 + \Delta\alpha_f/\alpha_f)\sim 18
\%$ - twice smaller than the respective change $ (\alpha_f
-\alpha_0)/\alpha_0 \sim 33\%$. Hence, statistics is unable to
explain this particular trend-like pattern in $\alpha$-exponent
values.\\
To see if it was a chance, we checked two other major crashes in
American market in 1987 and 1998. The scenario found for them is
revealed in Fig.5 and Fig.6. The '87 crash has also clear
decreasing trend in local $\alpha$ for the whole one year period
preceding the crash point (see details in Fig.5). At the same
time, DJIA signal constantly rises. These two contradicting trends
suggest that investors were becoming gradually more sceptic about
the market future at that time, despite the market index has been
rising.\\
The local $\alpha$ had started to reach the value below $0.45$
since the session $\sharp 25340$. Soon after, the DJIA index also
changed drastically its trend on 25.08.87 (session $\sharp
25385$). A comparison of $\alpha$ increase later on, with
simultaneous drop in DJIA signal between sessions $\sharp 25385$
and $\sharp 25395$, can be read as the confirmation of decreasing
trend in the market. It is worth to observe that the local
$\alpha$ gains again even deeper minimum $\alpha\simeq 0.43$ just
before the October crash. This minimum reveals again that the
market is very 'nervous' and that the increase of DJIA index,
initiated by session $\sharp 25403$, is only a small correction
before the forthcoming crash. The percentage drop
between maximal and minimal $\alpha$ value in the trend is about the same as in '29 crash. \\
The same phenomena occurs for '98 crash if we make its 'X-ray'
with the use of local $\alpha$-exponent as seen in Fig.6. The
market is much more nervous at that time (the DJIA index plot is
jagged much more than ever before). This leads to much smaller
values gained by $\alpha$ in its decreasing trend before
the crash.\\
In this particular case it is interesting to see what happened
next, i.e. after July 1998. The DJIA index run in the period
1995-2003 is displayed in Fig.7. We notice that just after the '98
crash, market entered into globally increasing, but very nervous
trend lasting up to January 2000. The $\alpha$-exponent increases
in this period, however, it probably gets many noise signals
coming from the frequently changing DJIA index value within
$N=215$ time-window length from the immediate past. In other words
the correlation range is too short with respect to $N$ and
$\alpha$-exponent loses its sensitivity to detect them. Therefore
the prediction of market signal evolution with the use of local
$\alpha$-exponent becomes difficult in the period 1995-2003.\\

\section{Conclusions}
We think that the careful analysis  of local $\alpha$-exponent
signal may significantly help to determine in which mode (strong
increasing, decreasing or unpredicted one) a market is at given
moment. As we have shown, the collected data suggest that very
clear trends can be distinguished not only for the financial
series signal but also for the local Hurst exponent. In
particular, we tested that $\alpha$ value drops significantly
before any crash in the DJIA signal is going to happen. This
confirms that local Hurst exponent may be used as a measure of
actual excitation state of the market. However, one has to
remember that strong, exterior phenomena not controlled by
investors are able to decelerate or accelerate predicted process,
drastically change its depth, or even its direction. Such sudden
exterior phenomena
can not, of course, be predicted by the local $\alpha$ behavior. \\
The best results in prediction of forthcoming drastic changes in
the market seem to be obtained if 'quiet' situation exists within
considered time-window length, i.e. if one has clear long-lasting
trends in market index before the crash. This situation took place
in 1929, 1987 and 1998. Otherwise, the $\alpha$-exponent contains
artifacts from the rapidly changing market index within
time-window over which it is calculated. In the latter case
predictions are troublesome.\\
We have checked that the above analysis, based on the local
$\alpha$-exponent, works well also for other market indices, like
the Warsaw Stock Exchange Index (WIG 20) [28], however different
time-window length should be used in that case. We believe that
observation and analysis of local Hurst exponent may play a
significant role as an additional hint for market investors, apart
from other technical indicators like moving average or momentum.
It could be a novel part of the classical technical analysis in
finances.

\subsection*{Acknowledgements}
One of the autor (D.G) would like to thank M. Ausloos for helpful
comments and discussion while this article was in preparation.

\newpage %1
\begin{figure}
\begin{center}
\includegraphics[width=14cm,angle=0]{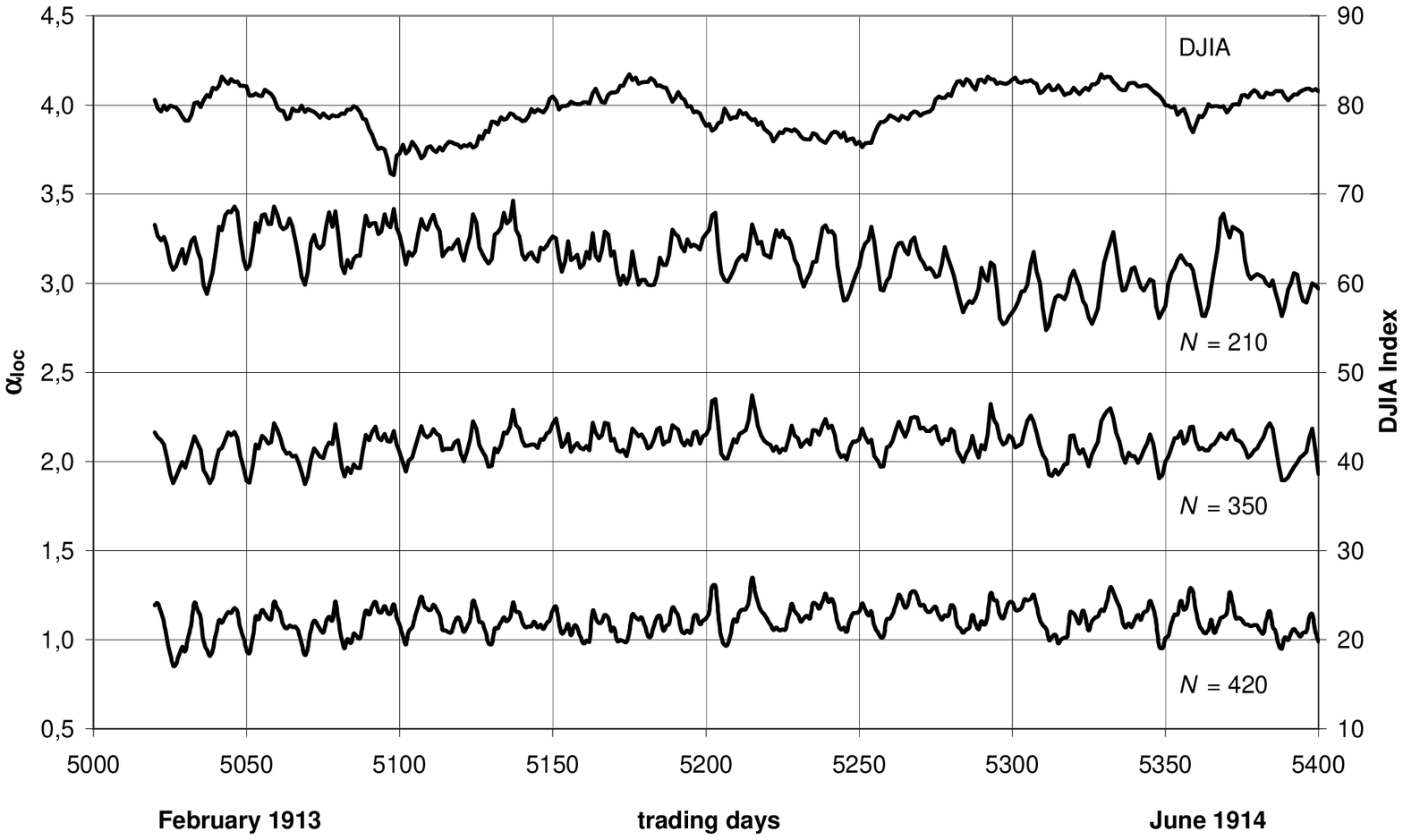}
\end{center}
 \caption{The example of time evolution plots of local $\alpha$-exponent for DJIA signal in the period
 Feb.1913--June 1914 for three different choices of time-window length $N$: $N=210$, $350$, $420$. The $\alpha$ values are artificially
 multiplied by two and then displaced along the vertical axis to make the differences noticeable.
 The DJIA index signal is also drawn for comparison.}
\end{figure}

\newpage    %2
\begin{figure}
\begin{center}
\includegraphics[width=14cm,angle=0]{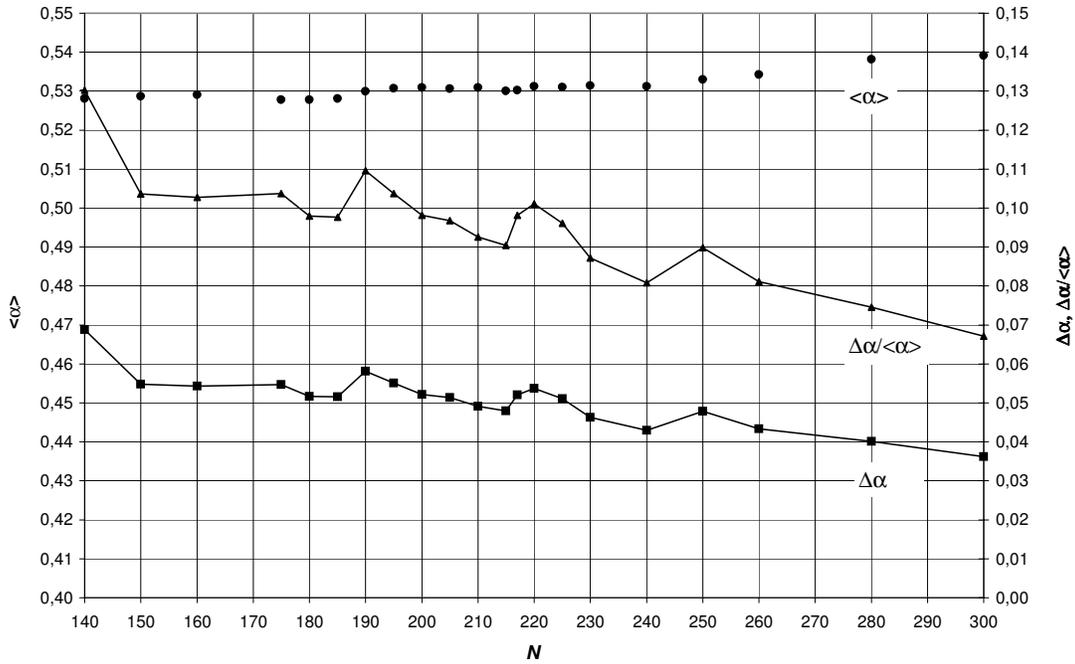}
\end{center}
 \caption{The standard deviation $\Delta \alpha(N)$ and the percentage uncertainty $\Delta\alpha/\langle\alpha\rangle$
 with respect to the mean $\alpha$ value $\langle \alpha\rangle$ as a function of time-window
 length $N$. Calculations were made on the sample of $500$ subseries of length $N$ each. Note
 that $\langle\alpha\rangle$, marked with round dots, stays steady and does not change significantly with $N$.
 It confirms the right choice of subseries in DJIA signal. Such subseries with uncorrelated signal
 and no drastic changes in $\alpha$-exponent (quiet domains) are good, natural tools to calibrate
 statistical uncertainties of the applied method.}
\end{figure}

\newpage %3
\begin{figure}
\begin{center}
\includegraphics[width=14cm,angle=0]{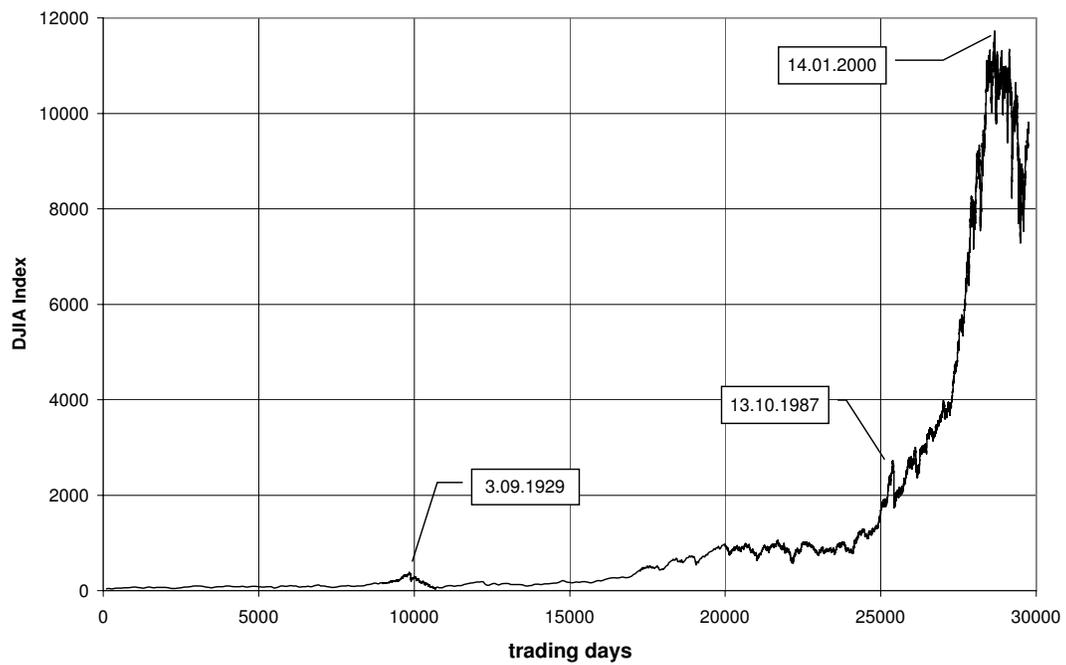}
\end{center}
 \caption{ The whole daily closure DJIA index history (1896-2003). The major economic events are seen as
 clear local maxima. They are marked with arrows indicating also the date of economic event.}
\end{figure}

\newpage    %4
\begin{figure}
\begin{center}
 \includegraphics[width=7cm,angle=0]{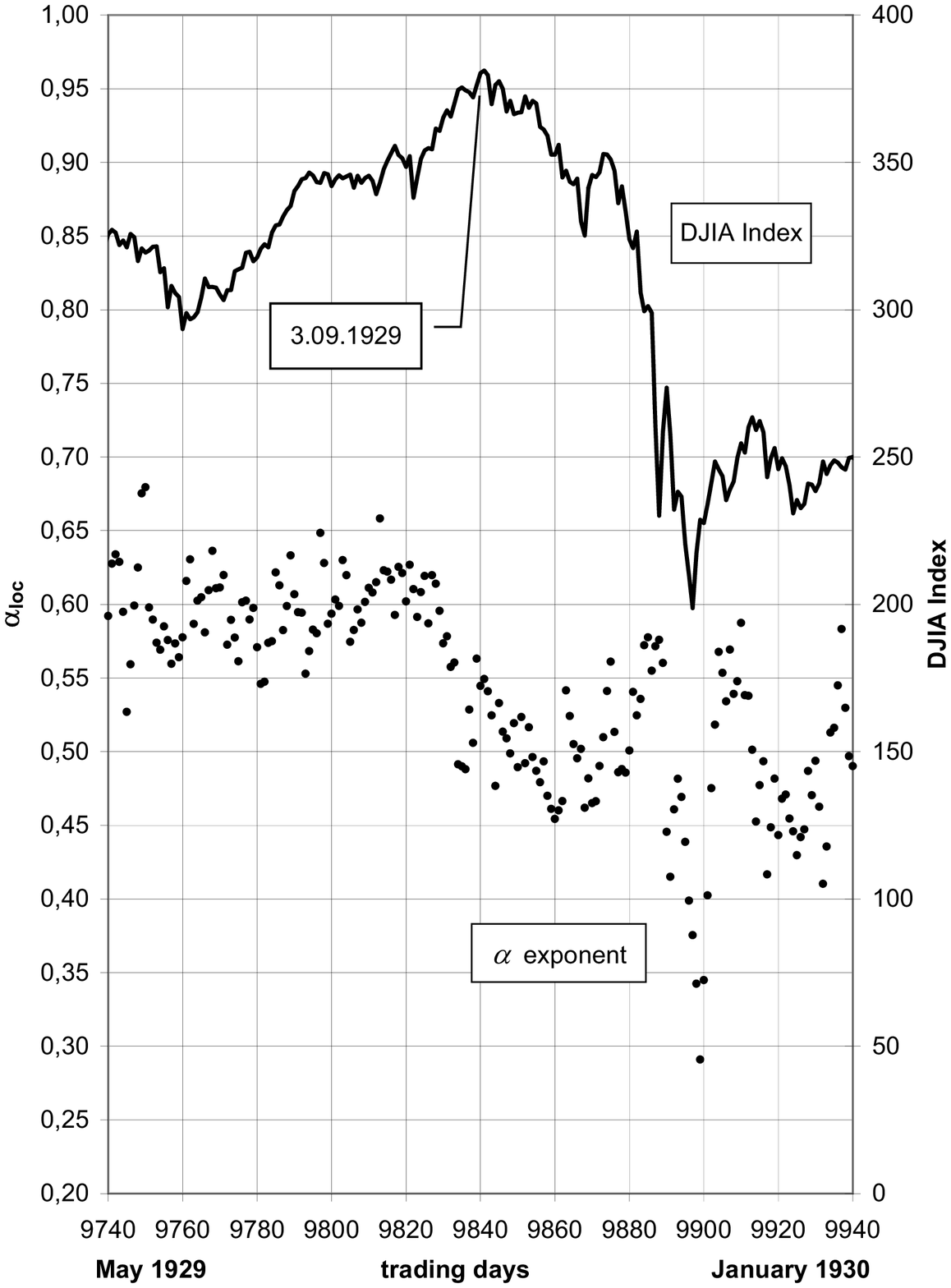}
  \includegraphics[width=7cm,angle=0]{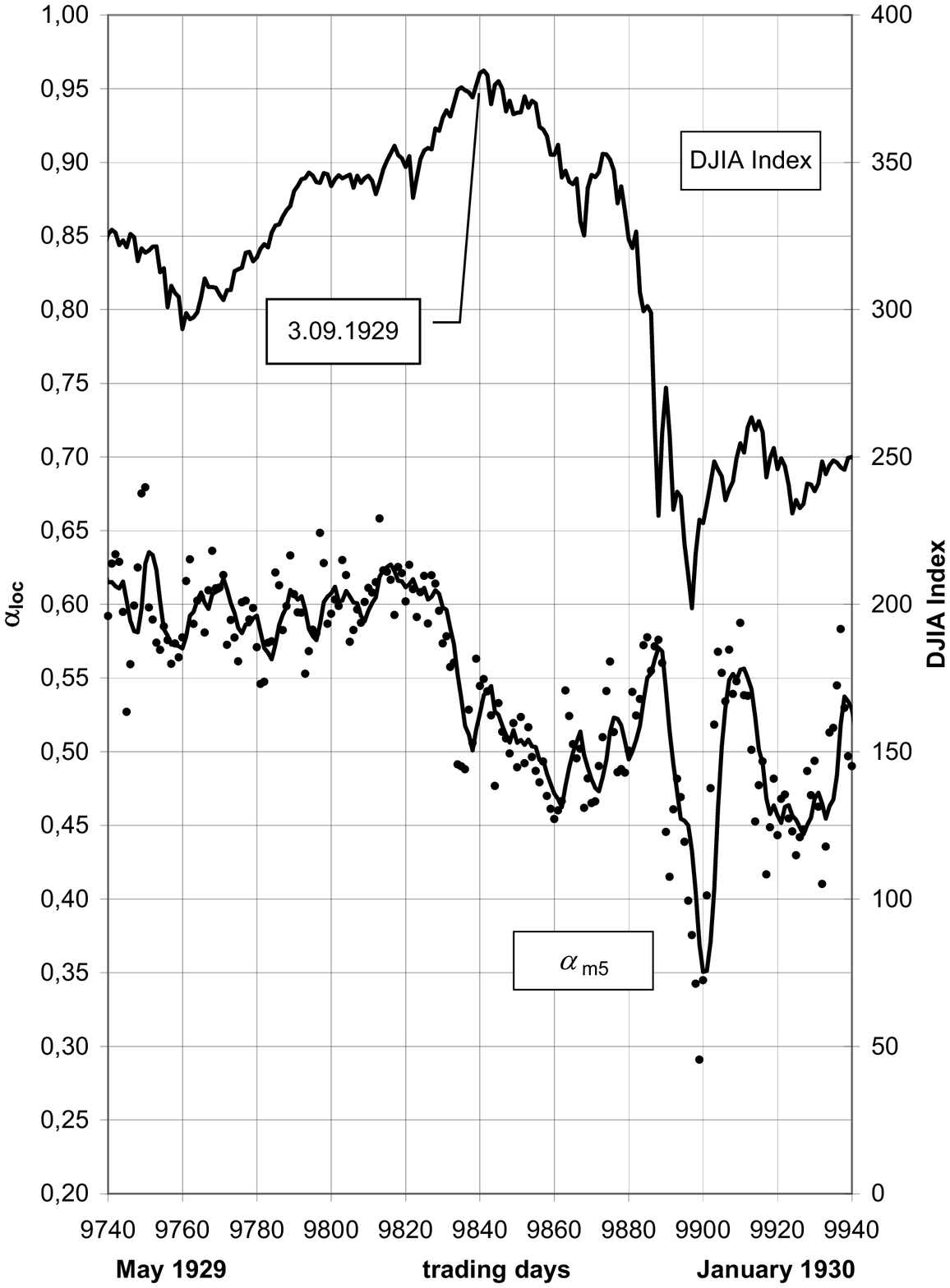}
\end{center}
 \caption{An 'X-ray' of the '29 crash made with the use of local
 $\alpha$-exponent.
\protect\newline
 (a) All local $\alpha$ values calculated session after session are marked as dots.
\protect\newline
 (b) The moving
 average $\alpha_{m5}$ of last five sessions.
 The significant drop in $\alpha$  between
 sessions $\sharp 9820$ and $\sharp 9860$ is striking.   }
\end{figure}

\newpage %5
\begin{figure}
\begin{center}
\includegraphics[width=7cm,angle=0]{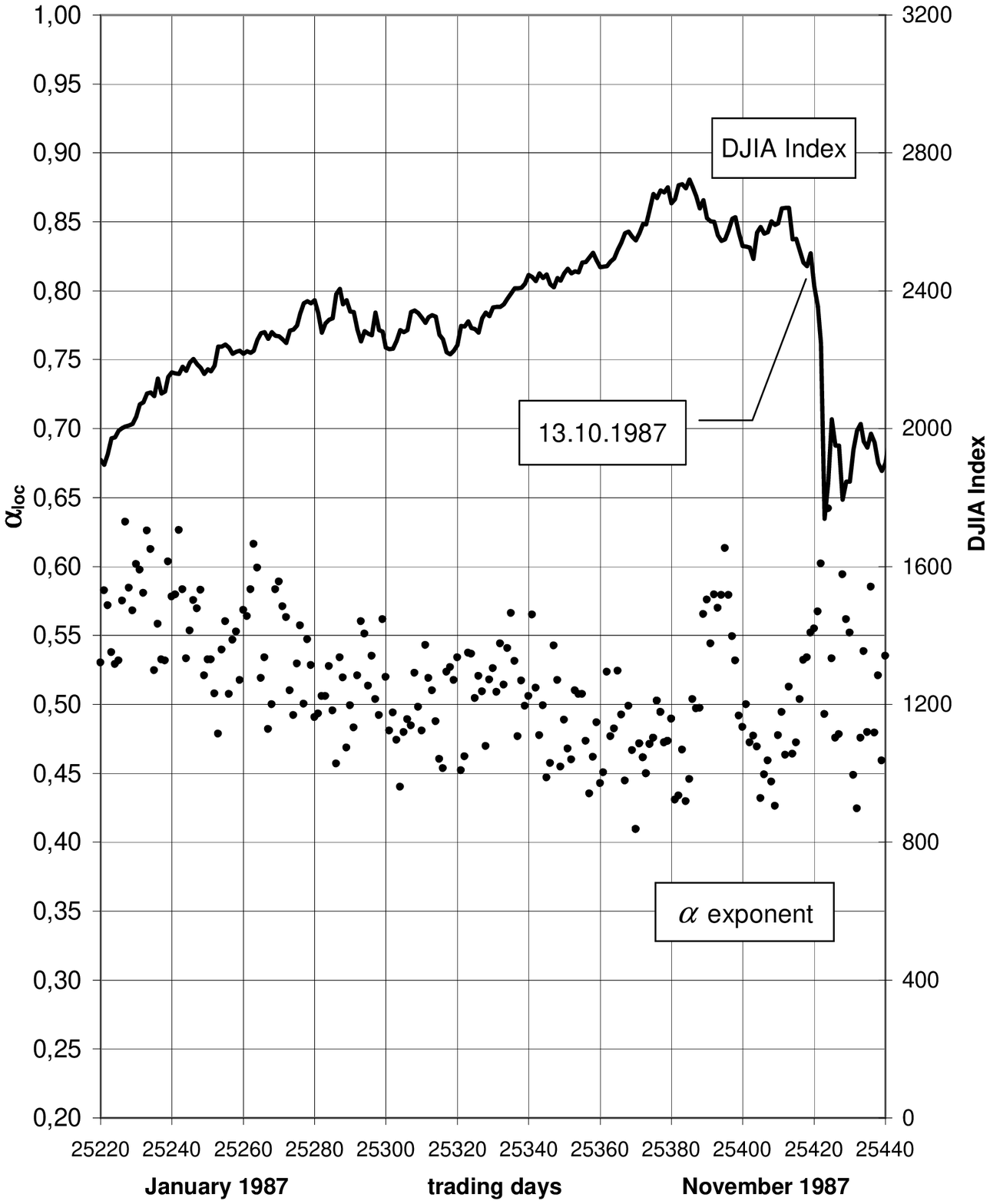}
 \includegraphics[width=7cm,angle=0]{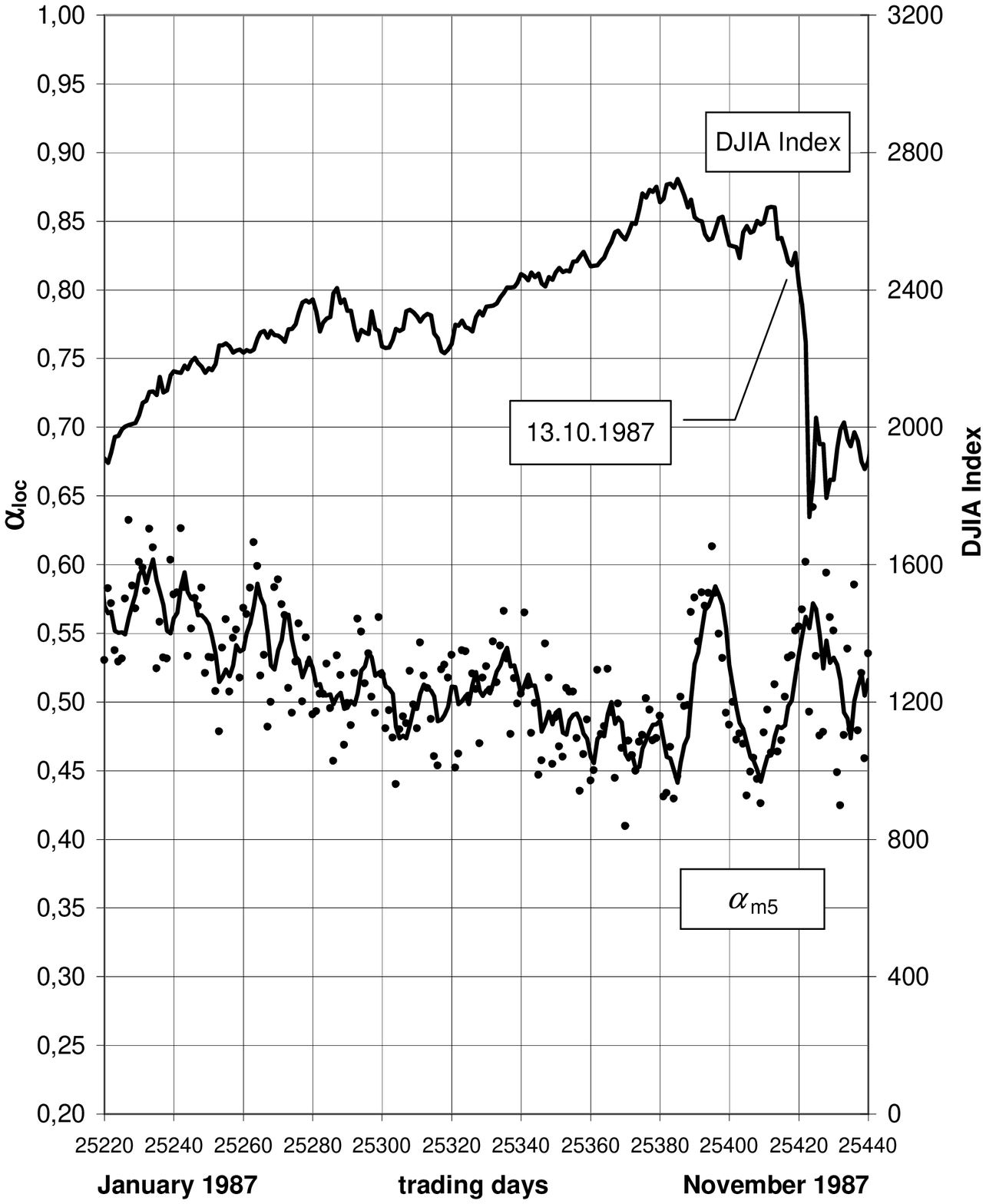}
\end{center}
 \caption{The history of '87 crash.
\protect\newline
 (a) The local $\alpha$ evolution calculated as before for $N=215$.
\protect\newline
 (b) The moving average $\alpha_{m5}$ line explains how the decreasing trend in $\alpha$ is formed.}
\end{figure}

\newpage %6
\begin{figure}
\begin{center}
\includegraphics[width=12cm,angle=0]{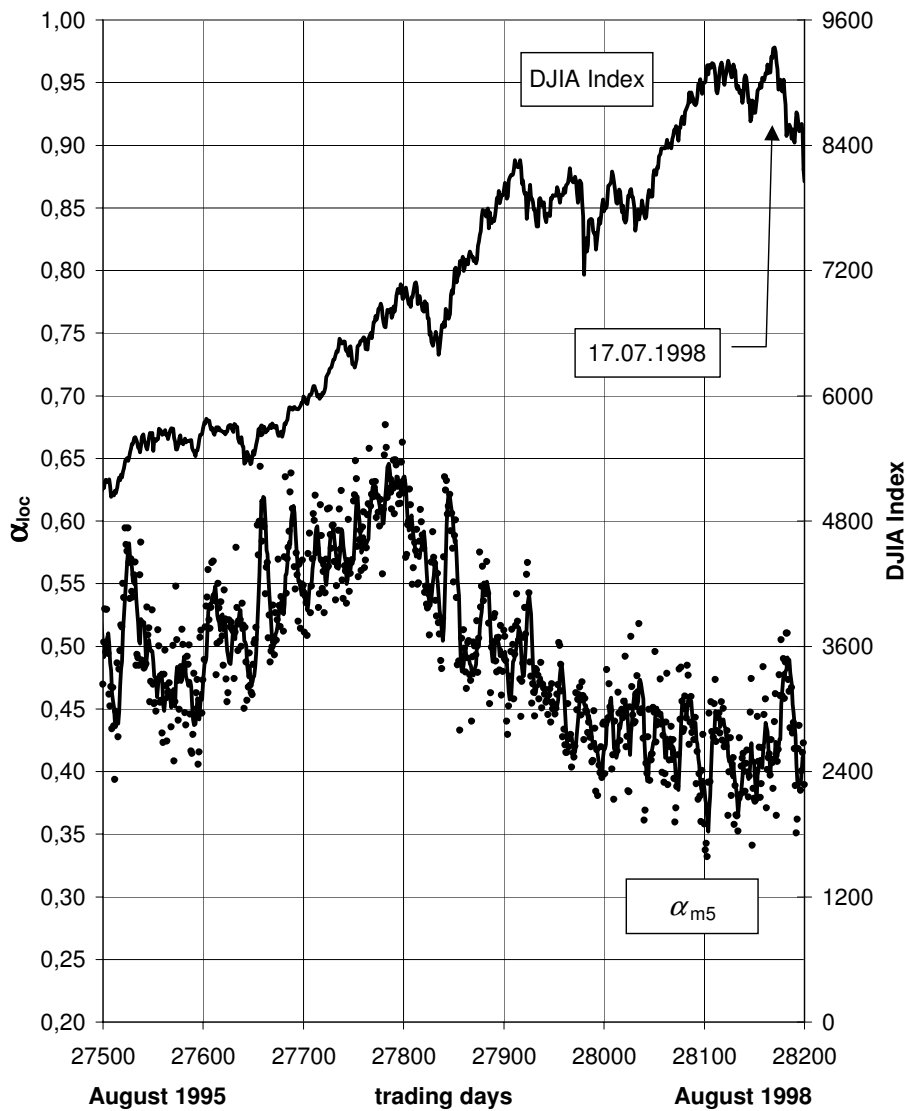}
\end{center}
 \caption{ The history of '98 crash. The same notation as previously applies.
\protect\newline
 The decreasing trend in $\alpha$ is much deeper ($50 \% $) than before. The investors seem to be
 very enthusiastic till the beginning of 1997
 ($\alpha > 0.65$), but they become
 pessimistic and nervous just one year later ($\alpha< 0.45$), despite the DJIA index still grows.
 The $\alpha$-exponent reaches its deep minimum ($\alpha<0.35$) in the half of 1998.
 This is a good moment for crash to appear(!).
}
\end{figure}

\newpage %7
\begin{figure}
\begin{center}
\includegraphics[width=12cm,angle=0]{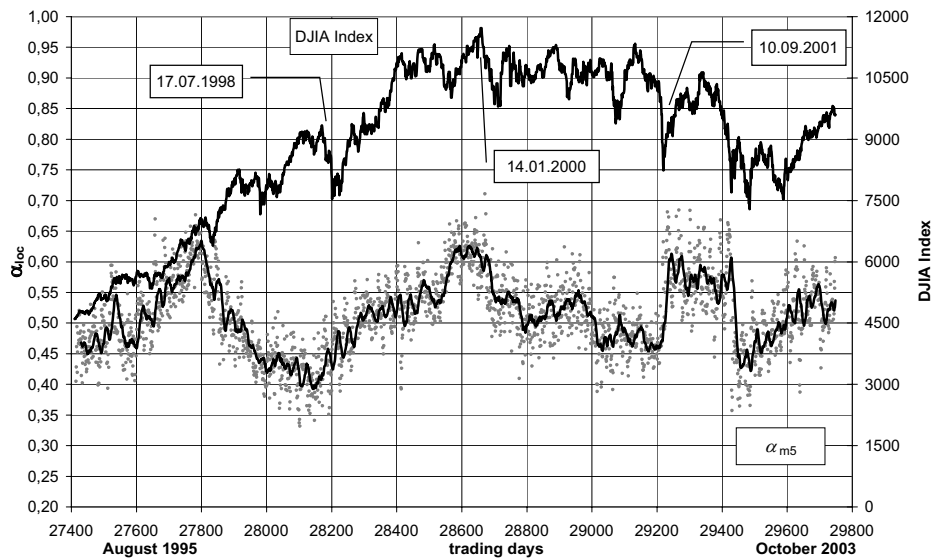}
\end{center}
 \caption{ The total display of the period 1995-2003 in DJIA signal history. The previous notation applies.
 Some important events are marked: the beginning of '98 crash (17.07.1998), the end of the longest
 economic boom in US history after $107$ months of its expansion (14.01.2000), terrorist attack in
 New York (11.09.2001).
 The decreasing trend in $\alpha$-exponent is observed from 14.01.2000 to 10.09.2001
 -- just one day before the terrorist attack. The huge drop in DJIA signal
 was then predicted at that time anyway (!). The DJIA index would have dropped
 immediately after, even if the attack did not take place (!).}
\end{figure}
\end{document}